\documentclass[a4paper, 10pt, conference]{ieeeconf}      

\makeatletter
\newcommand*\titleheader[1]{\gdef\@titleheader{#1}}
\AtBeginDocument{%
  \let\st@red@title\@title
  \def\@title{%
    \bgroup\normalfont\large\centering\@titleheader\par\egroup
    \vskip1.5em\st@red@title}
}
\makeatother

\IEEEoverridecommandlockouts                              

\usepackage{graphics} 
\usepackage{epsfig} 
\usepackage{fancyhdr}
\usepackage{amsmath} 
\usepackage{amssymb}  
\usepackage{xcolor}
\usepackage[colorlinks = true,
            linkcolor = black,
            urlcolor  = blue,
            citecolor = black,
            anchorcolor = black]{hyperref}
\usepackage{bm}
\usepackage{soul} 
\usepackage{subfigure}
\DeclareMathOperator{\Tr}{Tr}
\usepackage[
backend=biber,
style=numeric,
sorting=none
]{biblatex}
\title{\LARGE \bf Tire-road friction estimation and uncertainty assessment to improve electric aircraft braking system
}

\author{F.~Crocetti$^{1}$, 
             G.~Costante$^{1}$, 
             M.L.~Fravolini$^{1}$, 
             P.~Valigi~\IEEEmembership{Member,~IEEE}$^{1}$ 
\thanks{*This work was funded by Clean Sky 2 Joint Undertaking (CS2JU) under the European Union’s Horizon 2020 research and innovation programme, under project no. 821079, E-Brake}
\thanks{$^{1}$Department of Engineering, University of Perugia, Via G. Duranti, 93, Perugia, Italy. 
         {\tt\small {francesco.crocetti, gabriele.costante, mario.fravolini, paolo.valigi\}@unipg.it}}}%
}

\titleheader{This is an archival version of our paper. Please cite the published version \href{https://doi.org/10.1109/MED51440.2021.9480241}{https://doi.org/10.1109/MED51440.2021.9480241}
}
\begin{document}

\maketitle
\thispagestyle{empty}
\pagestyle{empty}

\begin{abstract}

The accurate online estimation of the road-friction coefficient is an essential feature for any advanced brake control system. In this study, a data-driven scheme based on a MLP Neural Net is proposed to estimate the optimum friction coefficient as a function of windowed slip-friction measurements. A stochastic NN weights drop-out mechanism is used to online estimate the confidence interval of the estimated best friction coefficient thus providing a characterization of the epistemic uncertainty associated to the NN block.  Open loop and closed loop simulations of the landing phase of an aircraft on an unknown surface are used to show the potentiality and efficacy of the proposed robust friction estimation approach.

\end{abstract}


 \section{INTRODUCTION}\label{sec:introduction}

The primary purpose of brake controllers is to enhance the braking efficiency maximizing the tire-road friction and preventing wheels from locking (ABS). Advanced anti-skid controllers take advantage of using the ElectroMechanically Actuated (EMA) braking systems that have found wide acceptance as a valid alternative to the conventional hydraulic ones \cite{bo2012research}. However, under the circumstances of sudden changes to the road conditions, a reliable estimation of the tire-road friction coefficient would lead to some relevant benefits in braking efficiency and passengers safety \cite{singh2012enhancement}. 
The estimation process is an extremely challenging and still open task due to the strongly nonlinear and uncertain physical phenomena. It is more evident in aeronautical operations due to the high-speed braking phase and the potential fast-changing runway.

The problem of estimating the tire-road friction coefficient has been extensively investigated.
In this paper, we focus on ``slip-oriented'' methods, which exploit the functional dependence of friction on slip (i.e. normalized difference between longitudinal and tangent velocities) to estimate the actual tire-road conditions. 
The Pacejka \cite{bakker1987tyre} and the Burckhardt models \cite{burckhardt1993,kiencke2005automotive} are widely used to characterize the normalised friction curve $\mu(\lambda)$ as a nonlinear function of the slip. This, in turn, allows the derivation of the longitudinal tire-force $F_{x}$ via the normal force $F_{z}$: $F_{x} := \mu(\lambda) F_{z}$. The Extended Kalman Filter has been used in \cite{ray1997nonlinear} to estimate a step-wise constant friction coefficient $\mu$, without any specific relation to the slip.
Later \cite{gustafsson1997slip},  \cite{muller2003estimation}  and \cite{baffet2007observer} proposed other simplified $(\lambda, \mu)$ models to estimate the actual road friction coefficient.  More recently, in \cite{tanelli2009real,tanelli2008real}, a least square and maximum likelihood approach has been proposed to estimate the parameters of a linearly parametrized approximation of the Burckhardt model, based on a sequence of $(\lambda, \mu)$ pairs as input, and using  a Quarter Car Model (QCM) for the system dynamics.
In \cite{de2011optimal} Recursive Least Square is used to online estimate the parameters of a similar linearized approximation of the Burckhardt model, and in \cite{de2012real} an enhanced constrained version of the same algorithm is proposed. 

Recently, machine learning techniques have been proposed for the estimation problem at hand.  In \cite{regolin2017svm} an SVM classifier has been used to improve the performance of a Burckhardt model-based EKF estimator.
 In \cite{zhang2017hierarchical} a general regression neural network (GRNN) is used to map the relationship between measured slip and the maximum friction coefficient.
Deep neural networks have been proposed in \cite{song2018estimating} to estimate the maximum friction value as a function of a large number of input signals, that may not be available in some brake control systems. 
In \cite{crocetti2020data}, the authors proposed an approach which, starting from a reduced number of measurements, uses a multi-layer neural network (MLP) to estimate, online, the best slip as a function of a windowed sequence of slip-friction pairs. The MLP training is based on a synthetic data-set derived from the Burckhardt model. 

Although the above mentioned neural networks methods are undoubtedly effective to map the uncertain and nonlinear relation between slip and friction, they do not provide any information on  the accuracy of the estimation. Precisely for this reason neural network estimators are very unlikely used within feedback control loops, especially in case of safety critical systems. To overcame this limitation, here we propose an approach, derived from \cite{Costante2020}, which gives a confidence interval for the  estimate. We believe that such an online accuracy information can be exploited within advanced control schemes.  As an example,  it may be possible to schedule different authority controllers as function of the estimated friction coefficient and its confidence interval.       




The first contribution of the study is given by a method, based on the idea in \cite{Costante2020}, providing an online estimation of epistemic uncertainty (real-time computation of a confidence measure) of the nonlinear neural network model of the slip-friction curve. The second contribution is the introduction of an improved synthetic data-set, and a MLP based on a reduced size input set with respect to the previous approach in \cite{crocetti2020data}, together with a more accurate dynamic model of the aircraft braking phase. 

The paper is organized as follows. Section \ref{sec:problem} formulates the problem of interest and discusses the underlying dynamic model. Section \ref{sec:estimation} presents the proposed estimation network and uncertainty computation, with the associated learning approach. Section  \ref{sec:simulations} discusses the MLP performance on a number of synthetic simulated scenarios. Finally, Section  \ref{sec:conclusions} draw some conclusions. 

\begin{figure*}[htb]
	\begin{center}
		\includegraphics[width=0.99\linewidth, height=4.1cm]{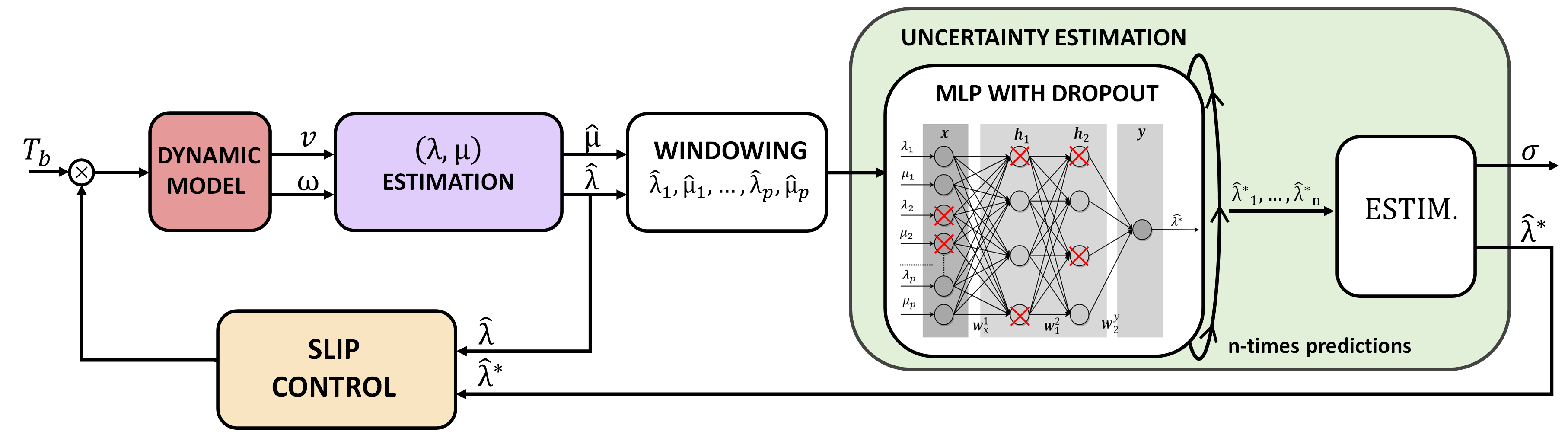} 
		\caption{The estimation and control scheme. The input vector is composed by sets of pairs $(\hat{\lambda}, \hat{\mu})$ obtained from the dynamic model, the measured ground speed $v$ and the angular velocity $\omega$. The uncertainty estimation block returns the MLP best slip estimate $\hat{\lambda^*}$ and the corresponding uncertainty $\sigma$. The predicted best slip value is also used as a set point for the feedback controller to regulate the requested braking torque $T_b$. Note that there is no differentiation between the left and right wheel for graphical representation simplicity. \vspace{-18 pt}}
		\label{fig:control_scheme}
	\end{center}
\end{figure*}
\section{Problem formulation and modelling} \label{sec:problem}

In this study we extend the classical Quarter Car Model (QCM) approach, taking into account also  relevant aerodynamic elements. A tricycle landing gear scheme is assumed for the wheel arrangement, such as the Piaggio P180 aircraft.

\subsection{Aircraft and friction modelling}
The system dynamics can be described by:
\begin{equation}
\begin{aligned}
M \dot{v} & =  - F_{x,L} - F_{x,R} - F_{D} \\  
J \dot{\omega}_{L} & =  r F_{x, L} - T_{b,L} \\
J \dot{\omega}_{R} & =  r F_{x, R} - T_{b,R} 
\end{aligned} \label{eq:aircraft_model}
\end{equation}
where $v$, $\omega_{L}$ and $\omega_{R}$ are the vehicle longitudinal speed and left and right wheel angular velocities, $J$ and $M$ are the mass and inertia moment, $r$ is the wheel radius, $T_{b,L}$ and $T_{b,R}$ are the left and right braking torque, $F_{x, L}$ and $F_{x, R}$ are the forces at the tire-road contact points, and $F_{D}$ is the drag force due to the wings, which is assumed proportional to the squared longitudinal velocity through a coefficient $K_{D}$.

The key elements in the model are the two forces $F_{x, L}$ and $F_{x, R}$, encoding the friction model of interest.
A widely adopted lumped model for $F_{x}$ assumes dependence on the vertical force $F_{z}$ acting at the tire-road contact point, on the longitudinal slip $\lambda$ according to the model:
\begin{equation}
\begin{aligned}
\lambda_{i} & := \dfrac{v - r \omega_{i}}{v}, \; \;
F_{x,i} = \mu_{i}(\lambda_{i}, \beta_{i}) \, F _{z},  \; \; i = L,R ,  
\end{aligned}
\label{eq:Friction}
\end{equation}
where the additional parameter vector $\beta$ characterizes the normalized friction function $\mu$ with respect to the actual runway surface. No dependence on slip angle is considered,  since the interest is for the aircraft landing phase. 
The model for the the vertical force $F_{z}$ is:
\begin{equation}
\begin{aligned}
F_{z} & =  Q_{g}\left(F_{g} - F_{P}\right) \\  
F_{g} & = M \, g \\ 
F_{P}  & = -K_{P} v^2
\end{aligned} \label{eq:verticalforces}
\end{equation}
where $F_{g}$ is the gravitational force and $F_{P}$ the lifting force, which is assumed proportional to the squared longitudinal velocity, and $Q_{g}$ is the coefficient modelling the distribution of weight on each of the two major wheels of a tricycle landing gear; we assumed here $Q_{g}=0.45$.

As for the function $\mu$, encoding the dependence of friction both on wheel slip and road characteristic, the static Burckhardt model \cite{burckhardt1993,kiencke2005automotive} is used:
\begin{equation}
\mu(\lambda, \beta) = \beta_{1} \left(1- e{\beta_{2}\lambda}\right) - \beta_{3}\lambda \, .
\label{eq:Burckardt_model} 
\end{equation}\emph{The Reference Road Scenarios}, i.e., the main road types we consider, are \emph{Asphalt dry}   $(\beta_{1} = 1.2801, \, \beta_{2} = 23.99, \, \beta_{3} =0.52)$, \emph{Asphalt wet}   $(\beta_{1} = 0.857, \, \beta_{2} = 33.822, \, \beta_{3} =0.347)$, and \emph{Snow}   $(\beta_{1} = 0.1946, \, \beta_{2} = 94.129, \, \beta_{3} =0.0646)$. The corresponding curves are depicted in the  Figure \ref{fig:Burckhardt}.

Let denote with $\mu^{*}$ the \emph{optimal friction}, i.e., the maximum of the friction curve, and with $\lambda^{*}$ the \emph{optimal slip}, i.e., the associated slip value. The presence of such a local maximum implies that, for each road type, there is a single slip value yielding the best braking performance. There are also other lumped models available in the literature, all of them exhibit such a single local maximum on friction.

\subsection{Problem formulation}
Based on the above model, the problem of interest in this paper is the development of data-oriented models for the estimation of the optimal slip $\lambda^*$ for the current runway type, and for the evaluation of the associated estimation uncertainty. 
The paper objective is the design of a multi-layer neural perceptron (MLP) and the associated learning strategy aimed at estimating the best slip  $\lambda^*$ based on a sequence of $(\lambda, \mu)$ pairs. In addition, a measure of estimation uncertainty is also required, which could allow for a subsequent control decision based on the estimation quality.

\section{THE FRICTION ESTIMATION NEURAL NET} \label{sec:estimation}
 
The overall estimation and control scheme is illustrated in Figure \ref{fig:control_scheme}. 
The measurements of the longitudinal and angular speed, and of the applied braking torque, are used to estimate the instantaneous slip and friction pair $(\lambda, \mu)$ based on the inversion of the dynamic model \eqref{eq:aircraft_model}. More sophisticated approaches could be used (see, e.g., \cite{tanelli2012combined}); this is left as an extension.  The scheme presented in this study is an improvement of the one presented in \cite{crocetti2020data}, where a simpler MLP structure and dynamic model have been considered, and the problem of prediction accuracy computation was not addressed.
The MLP training is based on a large number $N$ of input vectors $\boldsymbol{X_i}$, each one comprising $\boldsymbol{n}$ pairs of $(\lambda, \mu)$ . The idea is that a sliding window is used to collect a sequence of $\boldsymbol{n}$ pairs $(\lambda, \mu)$, and then fed to the MLP for estimation, according to the scheme in Fig. \ref{fig:control_scheme}.

Each vector $\boldsymbol{X_i}$ is associated to the corresponding true value $\lambda^*_{i}$ of the optimal slip (see figure 1). 
Overall, the training set $\mathcal{X}$ is given by the input-output data:
\begin{equation}
\mathcal{X} := \{(\boldsymbol{X_i}, \lambda^*_{i}), \;\; i=1, 2, \ldots, N\}.
\end{equation}
\subsection{The training dataset}
The crucial design elements of the data set $\mathcal{X}$ are the construction procedure of each windowed sequence $\boldsymbol{X_i}$, the windows size $\boldsymbol{n}$, and the   number $N$ of vectors in the training dataset.
The key idea is to use the Burckhardt model to generate a large number of friction curves by exploring the \emph{friction cube} according to the procedure proposed in \cite{crocetti2020data}. Each curve is further corrupted by AWG noise, to improve robustness and facilitate generalization, and both the original and the noisy curves have been used.
To extract the set of vectors $\boldsymbol{X_i}$, each curve has been sampled with 10.000 linearly spaced points. 
In addition, since the above discussed sampling approach generates ordered tuples, each  $\boldsymbol{X_i}$ is replaced by a shuffled copy. Overall, the dataset consists of $N=479320$ samples. The window size has been chosen equal to $n=15$, which provides the best RMSE error.   


\begin{figure}[thb]
\centering
		\includegraphics[width=0.99\linewidth]{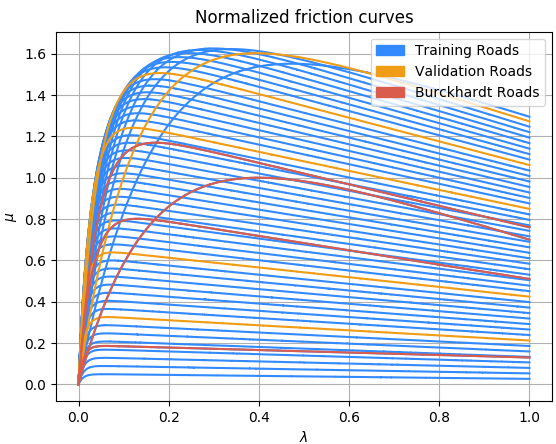} 
		\vspace{-20pt}
		\caption{The training dataset based on the Burckhardt model. The reference road conditions are shown in red color. Blue and yellow curves have been generated using different $(\beta 1,\beta 2,\beta3)$ values. \vspace{-10pt}}
		\label{fig:Burckhardt}

\end{figure}



\begin{figure*}[thb]
\centering
\subfigure[Uncertainty, slip and friction]{
\includegraphics[width=.45\textwidth]{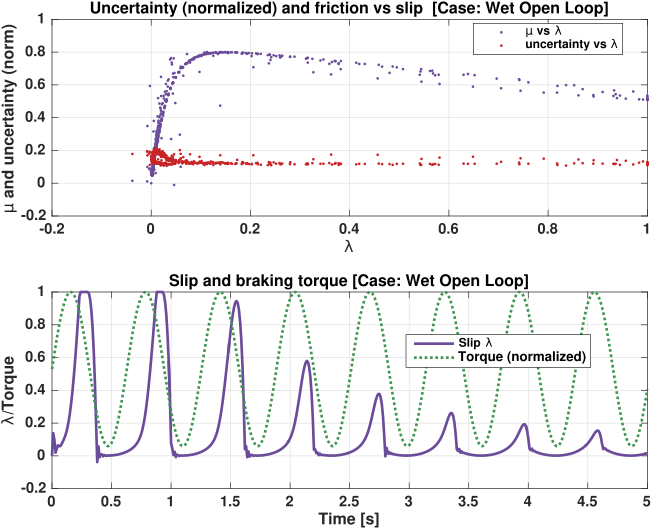}
\label{fig:openloopExp003a-Uncertainty}
}
\hfill
\subfigure[Best slip and uncertainty ]{
\includegraphics[width=.45\textwidth, height= 6.75cm]{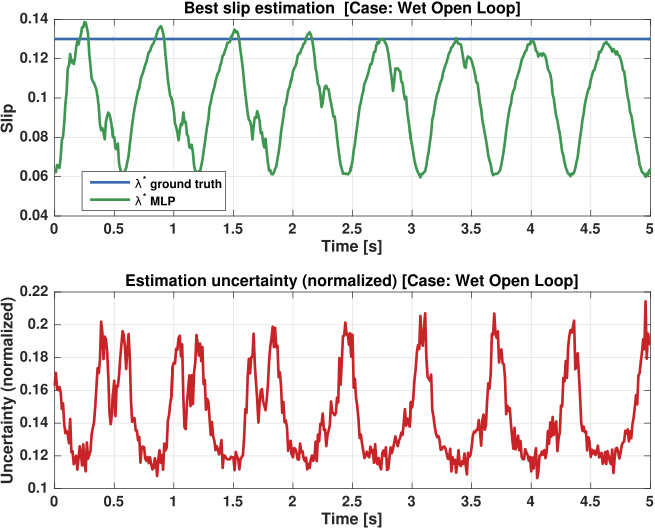}
\label{fig:openloopExp003a-prediction}
} 
\caption{Open loop experiment: a sinusoidal braking signal is applied in a wet road condition. The  $\lambda$, $\mu$ values \protect\subref{fig:openloopExp003a-Uncertainty} are used by the MLP to perform the estimates of the best slip value $\hat{\lambda^*}$ and to compute the corresponding uncertainty $\sigma$ \protect\subref{fig:openloopExp003a-prediction}.} \vspace{-10pt}
\label{fig:openloopExp003a}
\end{figure*}
\subsection{The Multilayer Perceptron Architecture}
As discussed in the previous sections, we rely on a multilayer perceptron model to estimate the best slip. The MLP processes the $15$ $(\lambda, \mu)$ pairs of the input vector $\boldsymbol{X_i}$ and outputs the estimated best slip value $\hat{\lambda}^{*}_{i}$.

The best performing architecture for our task is selected with the standard cross-validation technique and using the Mean Square Error (MSE) as the metric. The resulting configuration is composed of two hidden layers with $30$ neurons each and an output layer with a single neuron.
Each hidden layer (except for the last one) has a Rectified Linear Unit (ReLu) activation and is followed by a Dropout layer. The latter randomly "disconnects" links between hidden layers by sampling from a Bernoulli distribution parameterized by a hyper-parameter $p$ that encodes the probability to set to $0$ the corresponding input unit. The Dropout layer has a twofold role: during the training phase it acts as a regularizer to prevent overfitting \cite{srivastava2014dropout} while, at test time, it is used to compute the prediction uncertainty of the MLP model, as detailed in Section \ref{sec:unc_est}. As for the network hyper-parameters, the best value for $p$ is selected with cross-validation and set to $0.2$. 
The MLP is trained for 100 epochs using the Stochastic Gradient Descent (SGD) optimizer with a learning rate and weight decay set to 0.001 and 0.0001, respectively. The architecture and the model optimization are implemented with the Pytorch framework.

\subsection{Uncertainty estimation} \label{sec:unc_est}
Despite the outstanding representational capabilities of neural networks compared to other data-driven strategies, it is not straightforward to provide confidence intervals for their predictions. It is, indeed, uncommon to witness commercial products based on data-driven strategies that embed uncertainty estimation functionalities. However, this has been a very important research directions in the last decades \cite{gal2016uncertainty, khosravi2011comprehensive}.

To model prediction uncertainties, it is necessary to formalize the probability distribution of the output given the neural network inputs and its parameters. This is typically achieved by relying on a Bayesian formulation that allows to account for the \textit{epistemic uncertainty (EU)}, which models the uncertainty about the model parameters \cite{gal2016uncertainty}. More formally, to model the EU of the predicted best slip value $\hat{\lambda}^{*}_{t}$ associated to a new sample $\boldsymbol{X_t}$ containing $\boldsymbol{n}$ pairs of $(\lambda, \mu)$ it is necessary to compute the predictive posterior distribution $p(\hat{\lambda}^{*}_{t} \vert \boldsymbol{X_t})$. This is achieved by first introducing a prior distribution over the network parameters and a likelihood function. In our case, we choose to model both of them as Gaussian distributions: 
\begin{equation} \label{eq:eu_prior}
p(\mathbf{W}) \thicksim \mathcal{N}(0, l^{-2}\mathbf{I})
\end{equation}
\begin{equation} \label{eq:eu_lihelihood}
p(\mathbf{\hat{\lambda}^{*}_{t}} \vert f(\boldsymbol{X_t}, \mathbf{W})) \thicksim \mathcal{N}(f(\boldsymbol{X_t}, \mathbf{W}),\bm{\Sigma})
\end{equation}
where $\mathbf{W}$ encodes the neural network parameters, $l$ is a length-scale parameter that controls the prior belief over $\mathbf{W}$, $f(\boldsymbol{X_t}, \mathbf{W})$ is the neural network output and $\bm{\Sigma}$ represents the observation noise. Given those distributions and a supervised training set $\mathcal{X} = \{(\boldsymbol{X_0}, \lambda^{*}_{0}), (\boldsymbol{X_1}, \lambda^{*}_{1}), \dots, (\boldsymbol{X_N}, \lambda^{*}_{N})\}$, it is possible to compute the posterior distribution over the parameters $\mathbf{W}$ and, as a consequence, the predictive distribution as follows: 
\begin{equation}
p(\hat{\lambda}^{*}_{t} \vert \boldsymbol{X_t}, \mathcal{X}) = \int p(\hat{\lambda}^{*}_{t} \vert \boldsymbol{X_t}^{*}, \mathbf{W})p(\mathbf{W} \vert  \mathcal{X})d\mathbf{W} 
\end{equation}
where a marginalization over the network parameters has been performed.

For neural networks, the posterior distribution $p(\mathbf{W} \vert  \mathcal{X})$ is analytically intractable and, thus, we rely on approximate variational inference. With this strategy $p(\mathbf{W} \vert  \mathcal{X})$ can be approximated with a tractable distribution $q(\mathbf{W})$. In this work, this approximation is obtained by using dropout variational inference \cite{gal2016dropout}. In practice, it implies the minimization of the following objective function:
\begin{equation} \label{eq:loss_eu}
\small
\begin{split}
L_{EU} \propto \frac{1}{N} \sum_{i=1}^{N} & \left( \frac{1}{2} (\lambda^{*}_{i} - f(\boldsymbol{X_i}, \mathbf{\tilde{W}}_{i}))^T \bm{\Sigma}^{-1}(\lambda^{*}_{i} - f(\boldsymbol{X_i}, \mathbf{\tilde{W}}_{i})) \right. \\
& \left. +\frac{1}{2} \log\left|\bm{\Sigma}\right| \right) \\
&+ \sum_{j=1}^{L} \left( \frac{p_j l^2 \Tr(\bm{\Sigma})}{2N} \lVert \bm{M}_j\rVert^{2}_{2} + \frac{l^2 \Tr(\bm{\Sigma})}{2N} \lVert \bm{m}_j\rVert^{2}_{2} \right)
\end{split}
\end{equation}
where $L$ refers to the number of neural network layers, $\bm{M}_j$ and $\bm{m}_j$ indicate the variational parameter of the j-th layer and $p_j$ is the droput probability at the j-th layer. During the inference phase (\textit{i.e.,} after the optimization) the mean and the variance of the predictive distribution are computed through Monte Carlo integration by keeping active the dropout layers and performing $S$ stochastic forwards:

\begin{equation}\label{eqn:epistemic_mean}
\mathbb{E}_{EU}(\hat{\lambda}^{*}_{t}) \approx \frac{1}{S}\sum_{s=1}^{S}f(\boldsymbol{X_t}, \mathbf{\tilde{W}}_{s}) 
\end{equation}
\begin{equation} \label{eqn:epistemic_var}
\begin{split}
\mathrm{Var}_{EU}\left( \hat{\lambda}^{*}_{t} \right) \approx  \bm{\Sigma} & +  \frac{1}{S}\sum_{s=1}^{M} f(\boldsymbol{X_t}, \mathbf{\tilde{W}}_{s}) f(\boldsymbol{X_t}, \mathbf{\tilde{W}}_{s})^T \\ & - \mathbb{E}_{EU}(\hat{\lambda}^{*}_{t}) \mathbb{E}_{EU}(\hat{\lambda}^{*}_{t})^T
\end{split}
\end{equation}

In the experiments, at test time, we set $S$ to $500$.

\begin{figure*}[thb]
\centering
\subfigure[True and Estimated Best slip. Normalized Uncertainty]{
\includegraphics[width=.45\textwidth]{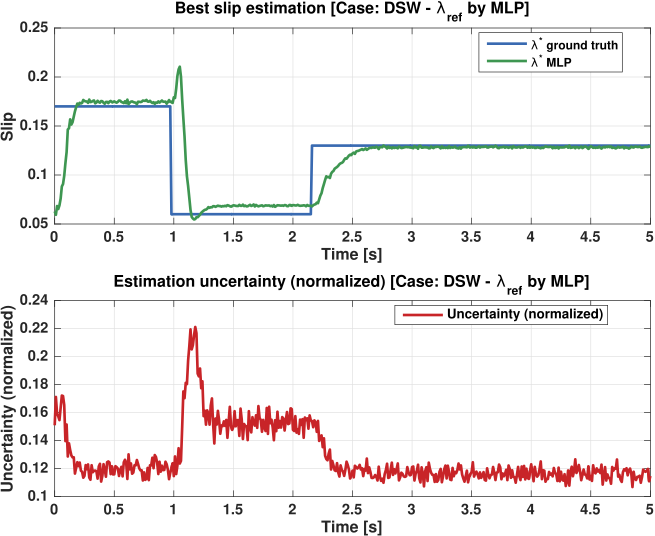}
\label{fig:predictionExp006b}
}
\hfill
\subfigure[Slip and Friction]{
\includegraphics[width=.45\textwidth]{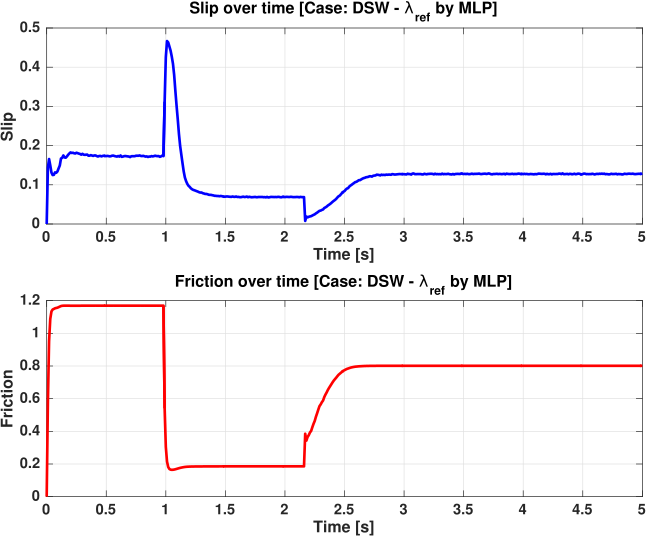}
\label{fig:lambdaMuExp006b}
}
\vfill
\subfigure[True and Estimated Best slip. Normalized Uncertainty]{
\includegraphics[width=.45\textwidth]{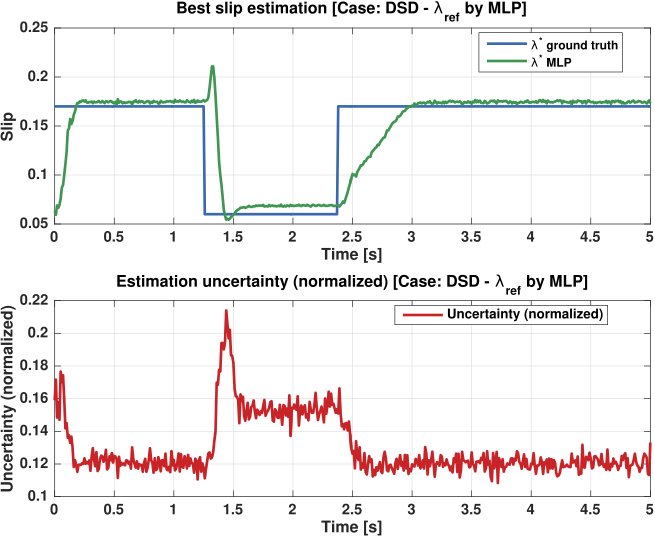}
\label{fig:predictionExp008b}
}
\hfill
\subfigure[Slip and Friction]{
\includegraphics[width=.45\textwidth]{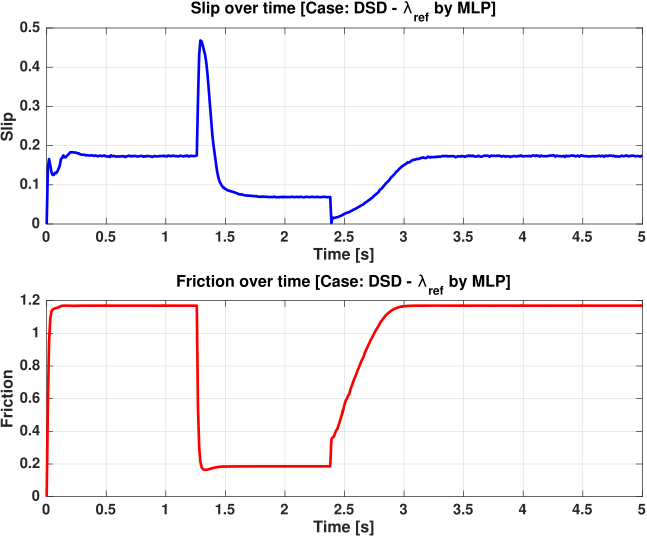}
\label{fig:lambdaMuExp008b}
}

\caption{Closed loop Experiments: two multi-transition road scenarios have been evaluated: $\text{Dry}\rightarrow\text{Snow}\rightarrow\text{Wet}$ (\protect\subref{fig:predictionExp006b}, \protect\subref{fig:lambdaMuExp006b}) and $\text{Dry}\rightarrow\text{Snow}\rightarrow\text{Dry}$ (\protect\subref{fig:predictionExp008b}, \protect\subref{fig:lambdaMuExp008b}). On the left side the online estimation of the MLP best slip and the normalized uncertainty estimates. On the right side, the time response of $\lambda$ and $\mu$. }
\label{fig:closedLoopSims}

\end{figure*}

\section{SIMULATION RESULTS} \label{sec:simulations}

The performances of the proposed MLP-based friction estimation algorithm have been evaluated through simulation studies, performed  both in open and closed loop conditions.
The reference Burckhardt roads in Dry, Wet and Snow conditions have been used to generate a number of challenging scenarios, either with fixed or with abrupt surface transition (see Table \ref{tab:roadConditions}). 
During the on line operation, the wheel experiences different skidding conditions and the vector of $(\lambda, \mu)$ pairs, computed by the estimation block (see Figure \ref{fig:control_scheme}) are used to feed the MLP estimation network. 
Simulations were performed assuming symmetrical operation for the left and right wheel. Without loss of generality, the following results will refer to the left wheel.

\begin{table}[t]
\centering
\resizebox{.35\textwidth}{!}{
    \begin{tabular}{|l|l|l|l|l|}
    \hline
    \multicolumn{5}{|c|}{\textbf{ROAD SCENARIO}}                                \\ \hline
    \multicolumn{1}{|c|}{\textbf{Fixed}} & \multicolumn{4}{c|}{\textbf{Multi-Transitions}} \\ \hline
    Snow  & SDS   & SWS  & SWD  &  DW  \\ \hline
    Wet   & WSW   & WDW  & WDS  &  WD    \\ \hline
    Dry   & DSD   & DWD   & DSW &  SD    \\ \hline
    \end{tabular}
}
\caption{\label{tab:roadConditions} Road conditions used to asses the uncertainties estimates provided by the algorithm}
\vspace{-20pt}
\end{table}

\subsection{Open loop case}

Results of a representative open-loop study are shown in Figure 
\ref{fig:openloopExp003a} where a feedforward sinusoidal braking torque with constant amplitude, bias, and frequency is applied, under the wet road scenario ($\lambda^* = 0.13$).
Such a braking torque drives the system to explore the entire slip interval (see Figure \ref{fig:openloopExp003a-Uncertainty} (lower)), hence the whole range of friction values are experienced during the simulation.
Figure \ref{fig:openloopExp003a-Uncertainty} (upper) shows the collected $(\lambda, \mu)$ pairs and the MLP normalized uncertainty.
The Figure shows that the uncertainty is larger in the low-slip region ($\lambda \in[0, 0.08]$). This enforce the idea that most of the roads in the low-slip region have similar $\mu(\lambda)$ response (see Figure \ref{fig:Burckhardt}); therefore, the MLP has inherent difficulties in discriminating the correct road condition. Viceversa, for larger slip values, differences among the curves increase, and therefore the MLP provides more accurate estimates, and the corresponding uncertainty decreases.
Figure \ref{fig:openloopExp003a-prediction} (upper) reports the corresponding time-response estimate $\hat{\lambda^*}$ provided by the MPL, while Figure \ref{fig:openloopExp003a-prediction} (lower) shows the corresponding uncertainty, normalized with respect to $\hat{\lambda^*}$.
It is observed that when the MLP estimate moves away from the  true optimal slip $\lambda^*$, the normalized uncertainty increases significantly (see Figure \ref{fig:openloopExp003a-prediction} (lower)), indicating that the actual friction prediction is less reliable. It is finally observed that the time intervals in which the prediction is good, can be easily identified considering only the periods when the uncertainty is minimum.

\subsection{Closed loop case \vspace{-4pt}}

The closed loop experiments have been performed by employing the  slip regulation Sliding Mode Controller proposed in \cite{de2013wheel} that modulates the braking torque required by the pilot.
In this scheme the best slip $\hat{\lambda^*}_{MLP}$ estimated by the MLP is used as the set point for the controller.
In figure \ref{fig:closedLoopSims} the closed loop simulation results for two Multi-transition road scenarios have been reported: $\text{Dry}\rightarrow\text{Snow}\rightarrow\text{Wet}$ (\ref{fig:predictionExp006b}, \ref{fig:lambdaMuExp006b}) and $\text{Dry}\rightarrow\text{Snow}\rightarrow\text{Dry}$  (\ref{fig:predictionExp008b}, \ref{fig:lambdaMuExp008b}).
In particular, figures \ref{fig:predictionExp006b} and \ref{fig:predictionExp008b} (upper) show the temporal evolution  of the MLP best slip prediction against the true values. Results indicate that the MLP tracks correctly the real best slip value and that the estimate uncertainty are influenced by the road type and the surface switching sequence. Moreover, the uncertainty significantly increases in case of transitions from high to low friction values.
This is further confirmed  by the $\lambda$ and $\mu$ time evolution shown in the figures \ref{fig:lambdaMuExp006b} and \ref{fig:lambdaMuExp006b}. It can be observed that when a transition toward a lower friction surface occurs, the applied braking force induces a sudden increase of the slip (thus moving in the right region of the Burckhardt curves). The sequence of ($\lambda$,$\mu$) pairs experienced during the transition phase, produces input vectors  that differ significantly from those used for training; nevertheless, such a pattern contains enough information to quickly infer the new road type.
Conversely, the transition to higher friction condition seems to be less critical due to the absence of uncertainty spikes following the transition. 


\section{CONCLUSIONS} \label{sec:conclusions}

In this paper a MLP has been proposed to provide an online robust estimation of the road friction coefficient of a landing aircraft on an unknown surface. A stochastic weights drop-out mechanism is proposed to online estimate the confidence interval for the estimation of the road best friction coefficient. Open loop and closed loop simulation experiments have shown that the method is effective in estimating the region of the $\lambda$-$\mu$ plane where the epistemic uncertainty produced by the NN block is large and therfore the online prediction in not fully reliable. It is expected that this information may be useful for the design of feedback controllers having different and selective control authority depending on the epistemic uncertainty level of the model.

\printbibliography

\addtolength{\textheight}{-12cm}   












\end{document}